# Structural investigation of (111) oriented $(BiFeO_3)_{(1-x)\Lambda}/(LaFeO_3)_{x\Lambda}$ superlattices by X-ray diffraction and Raman spectroscopy

J. Belhadi, S. Yousfi, H. Bouyanfif, M. El Marssi

LPMC EA2081, Université de Picardie Jules Verne, 33 Rue Saint Leu, 80000 Amiens, France

**Abstract**

$(BiFeO_3)_{(1-x)\Lambda}/(LaFeO_3)_{x\Lambda}$ superlattices (SLs) with varying $x$ have been grown by pulsed laser deposition on (111) oriented $SrTiO_3$ substrates. In order to obtain good epitaxy and flat samples a conducting $SrRuO_3$ buffer has been deposited prior to the superlattices to screen the polar mismatch for such (111) $SrTiO_3$ orientation. X-ray diffraction reciprocal space mapping on different family of planes were collected and evidenced a room temperature structural change at $x=0.5$ from a rhombohedral/monoclinic structure for rich $BiFeO_3$ to an orthorhombic symmetry for rich $LaFeO_3$. This symmetry change has been confirmed by Raman spectroscopy and demonstrates the different phase stability compared to similar SLs grown on (100) $SrTiO_3$. The strongly anisotropic strain and oxygen octahedral rotation/tilt system compatibility at the interfaces probably explain the orientation dependence of the phase stability in such superlattices.

# 1. Introduction

Multiferroic $ABO_3$ perovskite materials are currently the subject of intensive experimental and theoretical investigation motivated by their potential applications in optics, spin-tronics, multiple-state memories and sensors[1–3]. Bismuth ferrite ($BiFeO_3$ or BFO) is the most studied and promising multiferroic oxide thanks to its ferroelectric ($T_C$ ~1103 K) and G-type antiferromagnetic ($T_N$ ~640 K) orders at room temperature (RT)[4]. The coexistence and the coupling between these orders have prompted great interest to so-called MagnetoElectric (ME) effect in which a magnetic (electric) order can be controlled by an electric (magnetic) field[1,5]. Bulk $BiFeO_3$ crystallizes in a distorted rhombohedral symmetry with the R3c space group at RT with a pseudo-cubic lattice parameters $a_{pc(BFO)}$= 3.965 Å and $α_{pc}$=89.3–89.48°[4]. Its ferroelectric and magnetic properties are related respectively to the $Bi^{3+}$ lone pair ($6s^2$ orbital) and $Fe^{3+}$ ions. The promising piezoelectric properties and the large polarization which exceeds 100μC/$cm^2$ in (111) oriented films (the polarization is directed along [111]) make BFO thin films as an alternative to lead based materials for electromechanical applications[6]. In the last years, a large number of studies were performed on BFO thin films in order to modulate the structure and to improve the ferroelectric and magnetic properties by investigating the strain engineering, substrate orientation and chemical substitution[6–8]. For instance, rare earth substitution in Bismuth site ($Bi_{1-x}RE_xFeO_3$) (RE: Sm, Gd, Dy, La) leads to structural changes and to improved physical properties in thin films correlated to a morphotropic phase boundary (MPB) between a rhombohedral R3c phase and an orthorhombic Pnma phase[9,10]. Growing artificial superlattices (SLs) composed of epitaxial alternating layers of different materials on appropriate substrate is also a powerful way to tune the physical properties of the parent compounds by investigating the strain engineering and the interactions/coupling of the SLs layers[11–17]. This strategy can also lead to the possibility of creating a chemical- and strain-driven MPB for some composition in comparison to that observed for example in rare earth doped BiFeO$_3$ thin films[9,10]. Several theoretical and experimental studies were done on SLs based on the BiFeO$_3$ coupled with many others oxide materials such as ferroelectric BaTiO$_3$[18,19], PbTiO$_3$[20], Bi$_{0.5}$Na$_{0.5}$TiO$_3$[21], BiAlO$_3$[22], paraelectric SrTiO$_3$[23,24] and LaFeO$_3$[25–27], multiferroic BiMnO$_3$[28], antiferromagnetic BiCrO$_3$[29], magnetic and metallic La$_{0.7}$Sr$_{0.3}$MnO$_3$[30,31] and LaNiO$_3$[32], ferrimagnetic Fe$_3$O$_4$[6] and superconducting YBa$_2$Cu$_3$O$_7$[33]. However, the majority of the experimental studies concerning the multiferroic thin films and SLs were done on (100)-oriented substrates while a little effort has been concerned on the growth of these nanostructures on (111)-oriented substrates probably due to the polar nature of the atomic layers along [111] direction[34,35]. In the present work we investigate the effect of strain and substrate orientation on the structural interaction between BFO and LaFeO$_3$ (LFO) in (BiFeO$_3$)$_{(1-x)\Lambda}$/(LaFeO$_3$)$_{x\Lambda}$ [BFO$_{(1-x)\Lambda}$/LFO$_{x\Lambda}$] superlattices deposited on (111)SrTiO$_3$ substrate. The thickness of BFO layers (LFO layers) in SLs was varied, $0.2 \leq x \leq 0.8$, while the modulation period $\Lambda$ was kept constant at about 8nm. The superlattices and single films of BFO and LFO have been grown using pulsed laser deposition on (111) SrTiO$_3$ (STO) substrate buffered by a conducting layer of about 30 nm thickness of SrRuO$_3$ (SRO) and studied by x-ray diffraction and Raman scattering. Recently Carcan *et al.* revealed a nanoscale mixture in similar BFO/LFO SLs with $\Lambda$=8-9nm grown on (100) MgO and (100) STO substrates which was found to be strongly dependent on the BFO ratio in SLs (PbZrO$_3$ like versus Pnma like state)[27]. A PbZrO$_3$ antiferroelectric like state was discovered in BFO layers and a peculiar domain state was revealed in these SLs. For our knowledge no works have been done on BFO/LFO SLs grown on the polar STO (111) substrate and few works have been addressed on the (111) oriented BiFeO$_3$ based SLs[22,29]. Note that the (111) oriented BFO/LFO SLs is interesting because the polar axis of BFO is along (111) direction. In addition, the inter-planar spacing along (111) direction is smaller than the ones in (100) direction and thus can enhance the phase competition between layers in SLs at MPBs leading to nanostructures with enhanced properties. Strong interplay between oxygen octahedral at interfaces are indeed expected and could provoke new ordering as beautifully exemplified by the recent discovery of ferromagnetic ordering in LaFeO$_3$ and polar metals in LaNiO$_3$ for similar (111) orientation [36,37].

## 2. Experiments details

The samples (SLs and single films) were grown on oriented (111) STO substrate buffered with a conducting layer of SRO by pulsed laser deposition technique (MECA2000 chamber) using a KrF laser (248nm). The repetition rate and laser energy were fixed respectively at 4Hz and 1.5J/cm$^2$. BFO and LFO were grown under $5.10^{-2}$ mbar of oxygen pressure (PO$_2$) at 775°C. The SRO buffer layer was deposited on STO(111) substrate at 0.3mbar of PO$_2$ and 710°C both as a bottom electrode for future electrical characterizations but also in order to screen the polarity mismatch at the interface and to promote epitaxial growth of BFO and LFO layers. Structural characterizations (*ω/2θ*, rocking curve, reciprocal space maps and phi-scan) of the SLs and thin films were performed using a high-resolution 4-circles diffractometer with a Cu K$\lambda_1$ parallel beam (Bruker Discover D8). Raman spectroscopy measurements were performed using an argon ion laser (514.5 nm) and analyzed using a JobinYvon T64000 spectrometer equipped with a charge coupled device. The incident and scattered light (back-scattering geometry) was focused on samples using an objective x100 (spot of about 0.9 μm). Raman spectra were measured in both crossed Z(XY)$\bar{Z}$ and parallel Z(XX)$\bar{Z}$) geometries.

## 3. Results and Discussions

### *3.1. X-ray diffraction study*

Figure 1(a) shows a schematic of a BFO/LFO superlattice grown on (111)SRO(30nm)/(111)STO. Figure 1(b) shows the ω/2θ x-ray diffraction pattern of five BFO$_{(1-x)\Lambda}$/LFO$_{x\Lambda}$ SLs with *x*=0.2, 0.35, 0.50, 0.65 and 0.80 for a thickness of about 175-200nm and the ones of BFO(200nm) and LFO(200nm) single films. We note the presence of only *(111)* and *(222)* peaks for all superlattices and single films with no additional phase. The first order reflection of SRO buffer layer for SLs is not visible due to its small thickness and (111) peak position which is close to that of SLs. From the angular distance between the satellite peaks for SLs (star symbol in Fig.1*(b)*) we determined the periodicity Λ based on the modified Bragg formula of superlattice structures. The value of Λ was found between 7 and 8nm. The mosaicity of single films and SLs was determined using ω-scan (rocking curve). Figure 1*(c)* shows an example of ω-scan around (111) superlattice peak reflection for *x*=0.5 and for the (111) peak of the substrate. The FWHM value obtained for the superlattice is 0.26° compared to 0.1° for the (111) peak of single crystal substrate, indicating the good crystalline quality of the superlattice with low mosaicity. In order to investigate the in-plane epitaxial relationship between the SLs and the STO substrate a phi-scan investigation was carried out. Figure 1*(d)* shows a typical phi-scan obtained for the $BFO_{0.35\Lambda}/LFO_{0.65\Lambda}$ superlattice and STO substrate on the (020) family of plane. Three peaks separated by 120° are observed at the same angle for both the superlattice and the substrate. This result confirms the epitaxial in-plane lattice alignment between the superlattice and STO substrate. Note that same results were obtained on the other $BFO_{(1-x)\Lambda}/LFO_{x\Lambda}$ SLs, the SRO buffer layer and BFO and LFO single films.

From 2θ value of the SLs most intense satellite peak, we calculated the average out-of-plane interplanar spacing ($d^{SL}$) of all SLs using the Bragg formula. We plotted the results in Fig. 1*(e)* with the out-of-plane interplanar spacing of the BFO ($d^F_{BFO}$) and LFO ($d^F_{LFO}$) single films and the corresponding bulk parameters. The $d^F_{BFO}$ and $d^F_{LFO}$ are found to be 2.311Å and 2.301Å, respectively which are higher than their corresponding BFO and LFO bulk interplanar spacing parameters ($d^B_{BFO}$ and $d^B_{LFO}$, respectively). This is probably due to the in plane compressive strain applied by the substrate and/or to the presence of oxygen vacancies.

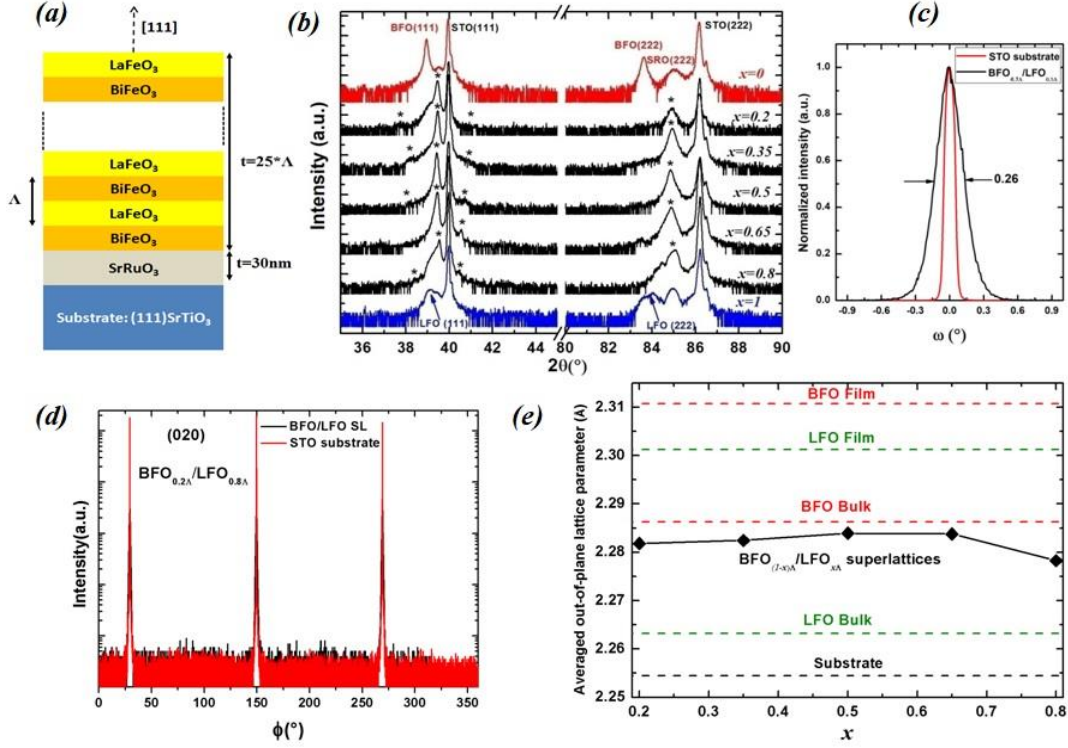

***Figure 1:*** *(a) Schematic of a BFO/LFO superlattice grown on (111)SRO(30nm)/(111)STO, (b) ω/2θ XRD patterns of $BFO_{(1-x)\Lambda}/LFO_{x\Lambda}$ SLs as a function of LFO ratio (0≤x≤1), (c) Rocking curve around (111) peak of $BFO_{0.5\Lambda}/LFO_{0.5\Lambda}$ superlattice and STO substrate, (d) phi-scan for the $BFO_{0.35\Lambda}/LFO_{0.65\Lambda}$ superlattice and STO substrate around (020) peak, (d) Averaged out-of-plane interplanar spacing of $BFO_{(1-x)\Lambda}/LFO_{x\Lambda}$ SLs as a function of LFO content. The out-of plane interplanar spacing of BFO and LFO films and bulk values are added for comparison.*

For the superlattices, the $d^{SL}$ present a small change as a function of LFO ratio in the period. The value of $d^{SL}$ is between $d^B_{BFO}$ and $d^B_{LFO}$ bulk parent compound. Thus, the results show a completely different behaviour compared to that observed for $Bi_{1-x}La_xFeO_3$ single films[9] and similar $BFO_{(1-x)\Lambda}/LFO_{x\Lambda}$ SLs grown on (100) STO and MgO substrate[27] in which a gradual decrease of the average out-of-plane lattice parameter with increasing the LFO ratio was reported. This later behaviour has been interpreted as a progressive diminution of the ferroelectric distortion when Bi is substituted by La. The results obtained here show clearly the role of (111) oriented substrate leading to different global strain state and layer interactions in (111) SLs compared to that obtained for the SLs grown on (100)STO[27].

To determine the in-plane interplanar spacing parameters and to reveal the eventual domain structures in our SLs and single films we have performed XRD reciprocal space maps (RSMs) around the *(h0l)* and *(hhl)* family of planes. We show in Fig.2 the results of RSMs around the (204) and (113) reflections for all SLs and around the (204) of BFO and LFO single films. For BFO and LFO single films (Fig.2(*a*) and (*b*) respectively) we remark the presence of only one reflection clearly shifted in contrast to the STO substrate reflection indicating a relaxation of the in-plane strain. However, in SLs the number of reflections depends on both the ratio of LFO *x* and RSM family of planes and the shift of these reflections with respect to STO is less pronounced compared to the ones of the single films indicating a partial strain relaxation from the substrate. Note that the any possible structural evolution in SLs when *x* varies is connected to (111) STO substrate orientation, interlayer interactions and symmetry compatibility between BFO and LFO in the SLs. It is important to precise that the global SLs intensity decreases on increasing *x* and that due to its structure factor, SRO diffraction peaks are more intense for even (204) reflection compared to odd (113) reflection. This explains the apparent increase of SRO contribution on increasing *x* in the (204) RSMs that is simply due to a change of intensity scale for better observing the SLs nodes.

Two reflections have been detected in (204) RSM (Fig.2*(c)*) for BFO-rich superlattice (*x*=0.2) while for (113) RSM (Fig.2*(d)*) three nodes can be distinguished. The third node ($Q_\perp$~7.22nm$^{-1}$ and $Q_\parallel$~4.15nm$^{-1}$) can either correspond to a satellite peak or to a domain. In this last case a monoclinic symmetry would also be a possible interpretation from three nodes in *(hhl)* RSM and two nodes in *(h0l)* RSM. For BFO$_{0.65\Lambda}$/LFO$_{0.35\Lambda}$ superlattice two reflections are observed for both (204) and (113) RSMs. When the ratio *x* of LFO increases in SLs, all these reflections in both (204) and (113) RSMs merge to constitute only one large reflection for the SLs with *x*>0.5. The presence of such splitting for BFO-rich SLs exclude the tetragonal symmetry and the evolution of the reflections when *x* increases indicates a change of the structure around *x*=0.50. A rhombohedral/monoclinic (resp. orthorhombic Pnma) like structure for SLs rich on (resp. LFO) BFO is therefore deduced from RSMs data. Note that the shape of RSM reflection for superlattice with *x*=0.8 is comparable to that observed in LFO thin film with a typical Pnma structure.

Figure 2*(e)* displays the evolution of the in-plane interplanar spacing ($d_{1-10}^{SL}$) of SLs calculated from (204) reflections as a function of the ratio *x*. The in-plane interplanar spacing of BFO ($d_{1-10}^{BFO}$) and LFO ($d_{1-10}^{LFO}$) thin films and bulk values of STO, BFO and LFO are also presented in Fig. 2*(e)*. For the SLs rich on BFO, two in-plane interplanar spacing are calculated. The differences is that the first one increases when *x* increases from a value close to the $d_{1-10}$ of the STO substrate and the second one is between BFO and LFO bulk values and decreases with increasing *x* and finally the two interplanar spacing merge at *x*=0.5. For *x*>0.5 the $d_{1-10}^{SL}$ increases with increasing the ratio of LFO in SLs. Therefore, a structural phase transition can be observed on increasing the LFO content in BFO$_{(1-x)\Lambda}$/LFO$_{x\Lambda}$ SLs. The same evolution of the in-plane lattice parameters obtained from (204) RSM for the BFO/LFO SLs grown on (100) STO was reported. On the other hand, for (113) RSM only one node was observed for the BFO/LFO (100) oriented SLs while a splitting of nodes is clearly observed for the (111) SLs rich on BFO. Such difference for the SLs with *x*<0.5 is probably related to two different structures. The Raman spectroscopy investigations presented in the next section will give more information about the nature of the phases in (111) SLs.

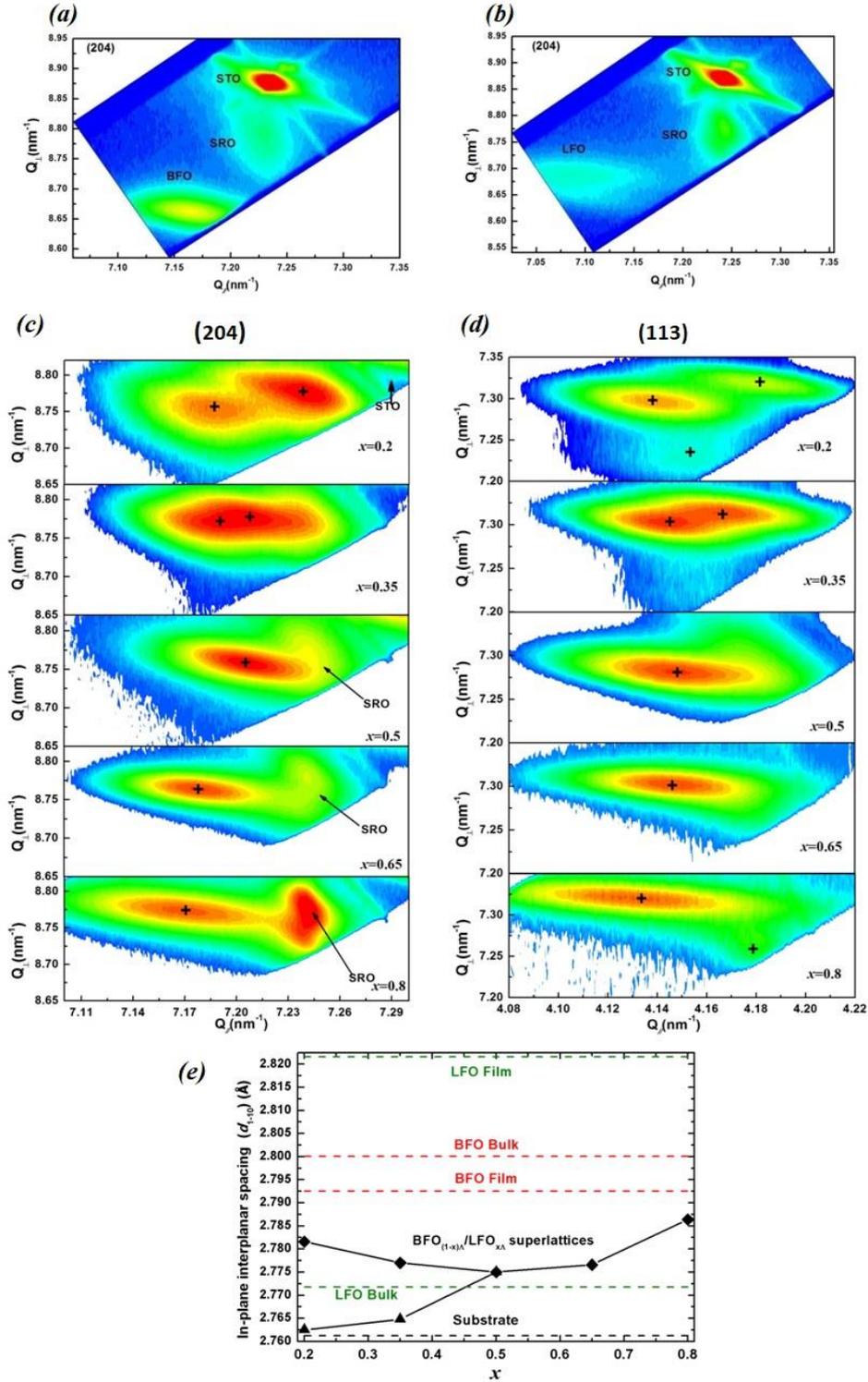

*Figure 2:* *reciprocal space map (RSM) around (204) for BFO (a) and LFO (b) films. (c) and (d) present RSMs of BFO$_{(1-x)Λ}$/LFO$_{xΛ}$ SLs (x=0.20, 0.35, 0.50, 0.65 and 0.80) around (204) and (113) respectively. (e) In-plane interplanar spacing ($d_{1-10}^{SL}$) of BFO$_{(1-x)Λ}$/LFO$_{xΛ}$ SLs as a function of LFO content. $d_{1-10}$ of BFO and LFO films and the $d_{1-10}$ bulk values are added for comparison.*

*3.2. Raman spectroscopy study*

In the rhombohedral symmetry (point group *R3c*), all optical modes $4A_1 + 9E$ of BFO should be observed by IR and Raman spectroscopy [38]. Moreover, the long-range electrostatic forces split all the $A_1$ and E modes into transverse optical (TO) and longitudinal optical (LO) components. The $A_1$ modes are polarized along Z axis parallel to the [111] polar axis and are allowed for the diagonal components of the Raman tensor while the doubly degenerate E modes are polarized along X and Y axis and can be observed in both parallel and crossed polarized geometries. Therefore the polarized Raman spectra can permit us to reveal the nature of phonon symmetries and assign the different modes.

The Raman spectra of (111) oriented BFO thin film and $BFO_{(1-x)\Lambda}/LFO_{x\Lambda}$ SLs are recorded in crossed $Z(XY)\bar{Z}$ and parallel $Z(XX)\bar{Z}$ normal backscattering geometries. Since the samples have an epitaxial single phase, we consider that the X and Y axes are perpendicular to Z axis and represent two perpendicular in-plane directions that are parallel to experimental stage (see Fig.3 (a)). The measurements are done on samples precisely aligned with respect to the crystallographic axes of the STO substrate: Z//[111] STO, Y//[11-2] STO, and X//[1-10] STO. Figure 3(b) displays the RT Raman spectra of BFO film recorded in crossed $Z(XY)\bar{Z}$ and parallel $Z(XX)\bar{Z}$ geometries.

The three modes at 135, 171 and 219 cm$^{-1}$ observed in $Z(XX)\bar{Z}$ geometry are assigned to $A_1$(TO) modes which are related to Bi atoms vibrations and oxygen octahedral tilt in the rhombohedral structure. The phonon observed at 135cm$^{-1}$ seems to be of $A_1$ symmetry according to Raman investigation on (111) oriented BFO crystal and single film [39,40]. However the assignment of the low frequency modes is debated in the literature since E mode appears close to this frequency[38,40,41]. The others weak modes observed in $Z(XX)\bar{Z}$ at 253, 312, 370, 424 and 518 cm$^{-1}$ are assigned to the E modes. These latter become more intense in the crossed $Z(XY)\bar{Z}$ geometries and are clearly identified. Our results are in agreement with earlier Raman investigation on single film and bulk[38–40,42].

Figures 3*(c)* and *(d)* display the room temperature polarized Raman spectra of $BFO_{(1-x)\Lambda}$/$LFO_{x\Lambda}$ SLs with $0.2<x<0.8$ recorded in crossed $Z(XY)\bar{Z}$ and parallel $Z(XX)\bar{Z}$ geometries, respectively. For the BFO-rich SL ($x=0.2$), two peaks appear at low frequency ($<200$ cm$^{-1}$) around 146cm$^{-1}$ and 171cm$^{-1}$ in the $Z(XX)\bar{Z}$ geometry and are reminiscent of the BFO $A_1$ phonon modes. The intensity of these two modes for this SL is relatively close while in BFO single film the intensity of the phonon at 171 cm$^{-1}$ is much higher than the low frequency mode that appears at 135 cm$^{-1}$. Note that these two phonons are characteristics of the R3c Rhombohedral polar state in BFO bulk and thin films and provide useful information about any possible symmetry changes. We remark a significant shift to high frequency of the first mode (146cm$^{-1}$ for $x=0.2$) compared to the (111) BFO film (130cm$^{-1}$) and its intensity decreases when $x$ increases and disappears completely for $x\geq0.65$. The frequency of the second mode for $x=0.2$ is close to that of the single BFO film (171cm$^{-1}$).When $x$ increases, the intensity of this mode decreases while its frequency increasesalmost linearly to reach 180 cm$^{-1}$ for $x=0.50$. These changes suggest a strong change in the Bi atomic displacements and the local structure in the SLs. In a recent work on $BFO_{(1-x)\Lambda}$/$LFO_{x\Lambda}$ SLs grown on (100)MgO substrate, this doublet of modes was observed at 150cm$^{-1}$ and 181cm$^{-1}$ for all SLs and was attributed to the change of polar ordering in the BFO layers from ferroelectric to antiferroelectric $PbZrO_3$-like state [27]. In our SLs the bands from the STO are large and no information could be gained in the frequency range above 200cm$^{-1}$. Separation of the different modes for identification is not feasible particularly for the LFO-rich SLs. For all SLs a strong phonon mode is observed at about 622cm$^{-1}$. The origin of this excitation is not clear and maybe related to a local disorder in the structure. For the SLs with $x\geq0.5$ this phonon corresponds to a doublet of mode and presents a LFO-like behaviour.

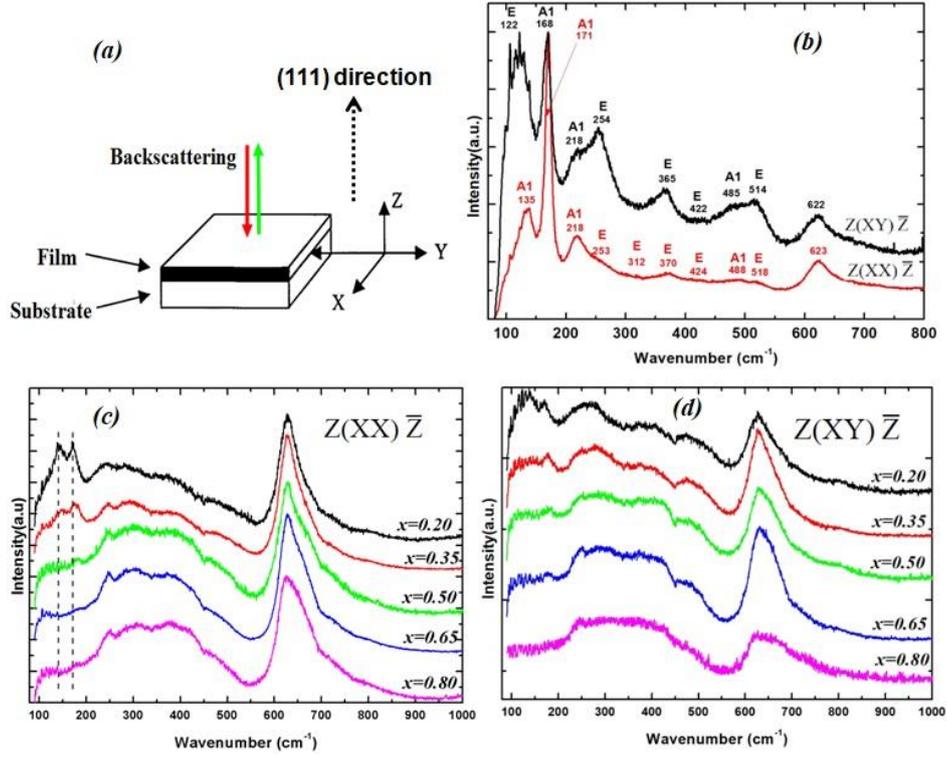

***Figure 3:*** *(a) schematic of backscattering geometry used for single films and SLs and (b) Room temperature polarised Raman spectra of BFO film in parallel and crossed geometries. (c) and (d) Room temperature polarised Raman spectra of $BFO_{(1-x)\Lambda}/LFO_{x\Lambda}$ SLs as a function of LFO ratio ($0.2 \leq x \leq 0.8$) recorded in parallel and crossed geometries, respectively. Dashed vertical lines in figure.3(c) are guide to the eyes and locate the 146 and 171 $cm^{-1}$ phonon bands of the x=0.2 SL.*

An antiferroelectric like $PbZrO_3$ state has been observed by Carcan *et al.* for rich BFO SLs on (001) oriented substrates while we observe here ferroelectric rhombohedral/monoclinic state on similar SLs grown on (111) oriented substrates[27]. This different phase stability is detected by both XRD and Raman spectroscopy. The key differences are observed on the number of nodes in (103) and (113) reciprocal space mapping of the (001) and (111) oriented rich BFO SLs. Another key difference is on the spectral Raman signature and the strong difference on the phonon frequency in the rich BFO SLs. While phonon reminiscent of BFO shows 170$cm^{-1}$ frequency for the rhombohedral/monoclinic SLs a strong hardening is observed for the antiferroelectric like SLs (shift to 181$cm^{-1}$) (see ref.27 and discussions therein).

The exact symmetry of the rich-BFO SLs is however not assigned. Indeed the XRD and Raman investigations combined together hint to symmetry different to the rhombohedral phase in the rich-BFO SLs. The XRD study rules out $PbZrO_3$ like structure but cannot distinguish between rhombohedral and monoclinic structure. On the other hand the two phonons observed by Raman spectroscopy and reminiscent of BFO present frequencies and relative intensities that are not in agreement with the R3c structure. The (111) orientation clearly modifies the phase stability of the BFO/LFO SLs compared to the (100) orientation and the key to understand the structural behaviour is probably based on the relative importance of the interlayer strain and oxygen octahedral tilt/rotation compatibilities. The interlayer strain seems identical to the (100) orientation with LFO (BFO) layers supposed to be under tensile (compressive) strain. However the influence of crystallographic orientation on the oxygen octahedral rotation differs totally as demonstrated by Moreau *et al.*[43] on (111) oriented $LaAlO_3$ with a similar rotation/tilt system compared to BFO. The authors demonstrated that the directions of oxygen octahedral rotations are neither parallel nor perpendicular to the surface of (111)-oriented film, in opposite to the strained (001) film in which the oxygen octahedral rotation axis is parallel or perpendicular to the surface. According to Moreau *et al.* compressive strain promotes monoclinic and triclinic phases while tensile strain stabilizes rhombohedral like structure. Interestingly, a monoclinic like structure may be present in the BFO layers under compressive strain in the (111) oriented rich-BFO SLs but the strong tendency of Bi off-centering and the (111) orientation would call in favour of the rhombohedral symmetry. Nano-twins are most likely present and the local structure may differ from the average structure as observed by XRD and only a refined TEM investigation would help decipher the complex structural interaction in this set of SLs.

## 4. Conclusion

BFO$_{(1-x)\Lambda}$/LFO$_{x\Lambda}$ superlattices have been grown by pulsed laser deposition on (111) oriented STO substrates. The polar mismatch inducing strong roughening has been screened by using an epitaxial conducting SRO buffer layer and good structural quality is obtained for the epitaxial SLs. Room temperature XRD RSMs on different family of planes and Raman spectroscopy show a structural phase change when the proportion of LFO in the period increases above $x$=0.5. BFO-rich SLs ($x$<0.5) are more likely of rhombohedral structure although a monoclinic distortion is possible while rich LFO SLs are of orthorhombic structure (Pnma). Phase stability on (111) oriented substrates differs compared to similar SLs deposited on (100) substrates. Strongly anisotropic strain and oxygen octahedral rotation/tilt system compatibility at heterointerfaces are probably responsible of this behaviour and a TEM investigation is under progress to unveil the local structural interaction.

**Acknowledgement**

This work has been funded by the region of Picardy (Project ZOOM) andEuropean Projects"NOTEDEV"FP7-People-ITN and H2020 RISE "ENGIMA n° 778072".